# Axisymmetric scattering of an acoustical Bessel beam by a rigid fixed spheroid

F.G. Mitri[*]

*Abstract* – Based on the partial-wave series expansion (PWSE) method in spherical coordinates, a formal analytical solution for the acoustic scattering of a zeroth-order Bessel acoustic beam centered on a rigid fixed (oblate or prolate) spheroid is provided. The unknown scattering coefficients of the spheroid are determined by solving a system of linear equations derived for the Neumann boundary condition. Numerical results for the modulus of the backscattered pressure ($\theta = \pi$) in the near-field and the backscattering form function in the far-field for both prolate and oblate spheroids are presented and discussed, with particular emphasis on the aspect ratio (i.e., the ratio of the major axis over the minor axis of the spheroid), the half-cone angle of the Bessel beam $\beta$, and the dimensionless frequency. The plots display periodic oscillations (versus the dimensionless frequency) due to the interference of specularly reflected waves in the backscattering direction with circumferential Franz' waves circumnavigating the surface of the spheroid in the surrounding fluid. Moreover, the 3D directivity patterns illustrate the near- and far-field axisymmetric scattering. Investigations in underwater acoustics, particle levitation, scattering, and the detection of submerged elongated objects and other related applications utilizing Bessel waves would benefit from the results of the present study.

## I. INTRODUCTION

BESSEL waves scattering by rigid (sound impenetrable) oblate and prolate spheroids is investigated using a formal solution based on the separation of variables in spherical coordinates, using the partial-wave (known also as normal-mode, or multipole) series expansion (PWSE) method, in contrast to the variational method [1], the direct numerical solution of integral equations [2], or the Helmholtz boundary integral representation of the incident and scattered fields (known as the *T*-matrix [3-5]). The advantage of the proposed method is the analogy with the partial-wave representation of the scattering by a spherical particle (or multiple spherical particles), introduced in [Section 334, pp. 272-282, [6]]. Moreover, for the case where the size of the spheroid is small compared to the wavelength of the incident radiation known as the Rayleigh limit, simplified expressions for the incident and scattered fields may be obtained and used to advantage for numerical predictions of the scattering.

*Corresponding author: F.G. Mitri (e-mail: F.G.Mitri@ieee.org).

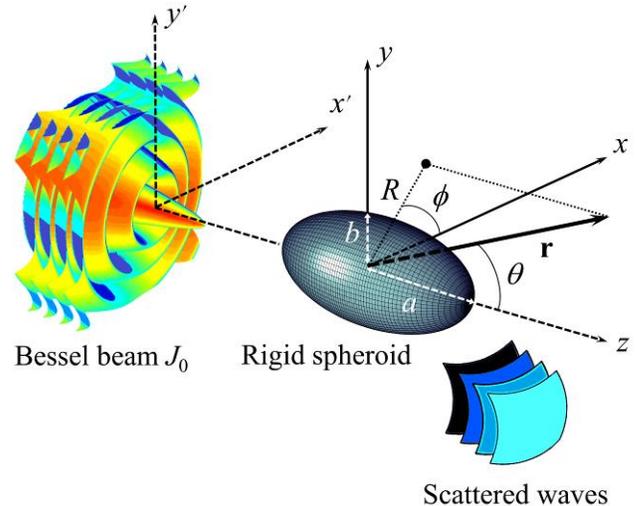

Fig. 1. The schematic describing the interaction of a monochromatic zeroth-order Bessel beam whose axis of wave propagation is centered on a rigid fixed (sound impenetrable) spheroid. The parameters $\theta$ and $\phi$ are the polar and azimuthal angles, respectively, in a spherical coordinates system.

Despite the recent analyses related to Bessel waves [7-11] (and *beams* generated from a finite aperture [12-15]) scattered by a *sphere*, analyzing the scattering by a *spheroid* using the theory formulated initially for a sphere as a first approximation, leads to significant inaccuracies, since there is substantial evidence that the scattering properties of elongated objects in plane waves differ quantitatively and qualitatively from those of a sphere. Moreover, many experiments revealed that acoustic levitation of (quasi)spherical liquid droplets (in air) changes their shapes to spheroids [16-18]. Note also that for liquid drops in air, there is a significant acoustic impedance mismatch at the air/liquid interface, and so liquid drops in air may be approximated as perfectly rigid. Thus, it is of some importance to develop an adequate formal solution to predict the scattering of Bessel waves by a rigid spheroid, which can be useful in various applications in underwater acoustics, particle manipulation and levitation, sonar imaging and other cases involving the interaction of spheroids with Bessel acoustical waves. A particular example includes the radiation force of progressive and (quasi-)standing Bessel waves [19] since it is intrinsically related to the scattering.

Although various works examined the acoustic scattering by a spheroid using the separation of variables method in spheroidal coordinates [20-24], other analyses have investigated the *T*-matrix (or null-field) method [4, 5, 25-29] the finite element method (FEM) [30], the finite difference [2] time-domain method [31], the fast multipole [32] accelerated boundary element method (BEM) [33], the 3D-BEM [34], the 3D-infinite element method based on a prolate spheroidal multipole expansion [35], and the shape perturbation method [36] to name a few, with each of these methods having their own associated advantages, disadvantages, and conditions of applicability [26].

In this analysis, the incident and scattered pressure fields for Bessel waves are directly expressed in terms of PWSEs, and the Neumann boundary condition (for a rigid spheroid) is applied to evaluate the scattering coefficients. Once these coefficients are obtained, computation of the scattering becomes possible anywhere (i.e., near- or far-field regions) in the fluid space surrounding the spheroid. In Section II, the theoretical formalism for the scattering (not taking into consideration the contribution of evanescent waves known to decay rapidly away from the fluid-scatterer interface) is developed, while in Section III, numerical computations and discussions illustrate the theory. Section IV provides a summary of this work.

## II. THEORETICAL FORMALISM

Consider an acoustical monochromatic zeroth-order Bessel beam propagating in a nonviscous fluid, and incident upon a spheroid centered on its axis of wave propagation [i.e., end-on incidence (Fig. 1)]. In a system of spherical coordinates ($r$, $\theta$, $\phi$) with its origin chosen at the center of the spheroid (Fig. 1), the incident pressure field is expressed as a PWSE as [7, 37],

$$P_i(r,\theta) = P_0 J_0(kr\sin\theta\sin\beta) e^{i(kz\cos\theta\cos\beta - \omega t)}$$
$$= P_0 e^{-i\omega t} \sum_{n=0}^{\infty} i^n (2n+1) j_n(kr) P_n(\cos\theta) P_n(\cos\beta), \quad (1)$$

where $P_0$ is the pressure amplitude in the absence of the waves, $J_0(.)$ is the cylindrical Bessel function of zeroth-order of the first kind, $j_n(.)$ is the spherical Bessel function of the first kind, $P_n(.)$ are the Legendre functions, $k$ is the wave number, and $\beta$ is the half-cone angle of the Bessel beam.

Upon the interaction of the acoustical waves with the spheroid, the beam is scattered, and is mathematically represented by a scattered complex pressure field expresses as,

$$P_s(r,\theta) = P_0 e^{-i\omega t}$$
$$\times \sum_{n=0}^{\infty} i^n (2n+1) s_n h_n^{(1)}(kr) P_n(\cos\theta) P_n(\cos\beta), \quad (2)$$

where $s_n$ are the spheroid's scattering coefficient to be determined by applying appropriate boundary conditions with the assumption that the fluid is nonviscous and the object is perfectly rigid, and $h_n^{(1)}(.)$ is the spherical Hankel function of the first kind.

The use of the PWSEs to describe the incident and scattered fields by multipole expansions of spherical waves (chosen as the basis functions), is justified since they generally form a complete set not only on spherical surfaces, but also on non-spherical surfaces (pp. 169-190 in [5]). This is an early result discussed by several authors [4, 38, 39] from the standpoint of wave diffraction and scattering theories.

The surface shape function $S(\theta)$ of the spheroid has an azimuthal symmetry when centered on the symmetric Bessel beam, thus, it only depends on the polar angle $\theta$. Its expression is given by,

$$S(\theta) = \left(\cos^2\theta/a^2 + \sin^2\theta/b^2\right)^{-1/2}. \quad (3)$$

The parameter $b$ is the equatorial radius of the spheroid, and $a$ is the distance from the center to the pole along the symmetry axis $z$. An *oblate* spheroid is defined such that $a < b$ (cf. Fig. 1), whereas a prolate spheroid is defined such that $a > b$. It is obvious that when $a = b$, the surface shape function $S(\theta)$ no longer depends on the polar angle $\theta$, which corresponds to the surface shape function of a sphere.

Note that the domain of applicability of the so-called "Rayleigh hypothesis" [40, 41], which assumes adequate convergence of the PWSEs up to the object's boundary owing to Huygens' principle [42], and is usually valid for an ellipse/spheroid with low aspect ratio, i.e. $a/b < \sqrt{2}$ [43], can be extended to spheroids of larger aspect ratio by increasing the maximum value of $n$ in the PWSEs required for convergence [44]. This has been also utilized to evaluate the acoustic radiation force on rigid prolate and oblate spheroids in Bessel beams composed of progressive, standing and quasi-standing waves [19]. Nonetheless, this maximum value of $n$ cannot in practice increase indefinitely because of an eventual ill-conditioning susceptible to the system of linear equations (see Section III).

Therefore, applying the Neumann's boundary condition for the total (incident + scattered) steady-state (time-independent) pressure field for a rigid immovable spheroid at $r = S(\theta)$, leads to the boundary condition

$$\nabla(P_i + P_s) \cdot \mathbf{n}\big|_{r=S(\theta)} = 0, \quad (4)$$

where,

$$\mathbf{n} = \mathbf{e}_r - \left([1/S(\theta)]dS(\theta)/d\theta\right)\mathbf{e}_\theta, \quad (5)$$

with $\mathbf{e}_r$ and $\mathbf{e}_\theta$ denoting the unit vectors along the radial and polar directions, respectively.

The scattering coefficients $s_n$ for the spheroid can be now be determined after substituting (1) and (2) into (4) using (5). This procedure leads to a system of linear equations,

$$\sum_{n=0}^{\infty} i^n (2n+1) P_n(\cos\beta)\left[\Gamma_n(\theta) + s_n \Lambda_n(\theta)\right] = 0, \quad (6)$$

where the functions $\Gamma_n(\theta)$ and $\Lambda_n(\theta)$ are expressed, respectively, as

$$\begin{Bmatrix} \Gamma_n(\theta) \\ \Lambda_n(\theta) \end{Bmatrix} = \begin{Bmatrix} j_n'(kS(\theta)) \\ h_n^{(1)'}(kS(\theta)) \end{Bmatrix} P_n(\cos\theta) k$$
$$+ \begin{Bmatrix} j_n(kS(\theta)) \\ h_n^{(1)}(kS(\theta)) \end{Bmatrix} \frac{1}{S(\theta)^2} \frac{dS(\theta)}{d\theta} P_n^1(\cos\theta). \quad (7)$$

The primes denote a derivative with respect to the argument,

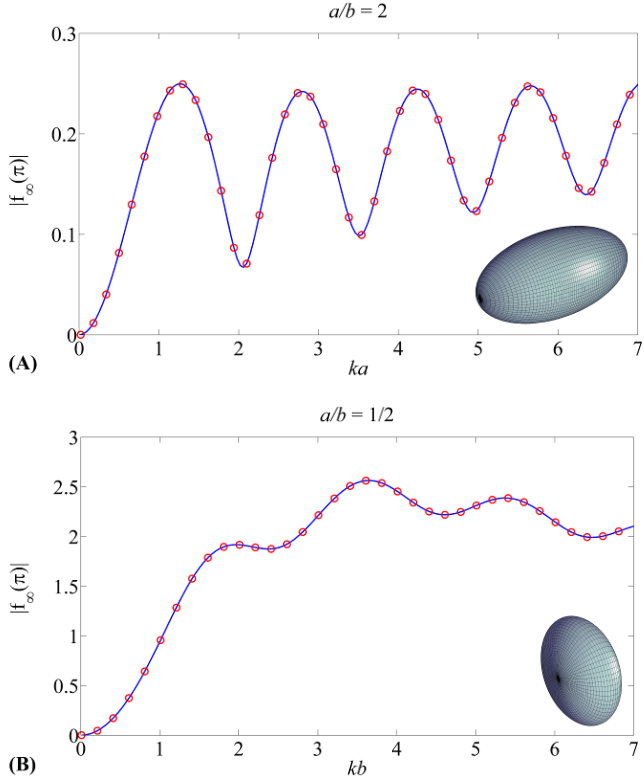

Fig. 2. The magnitude of the far-field backscattering ($\theta = \pi$) form function for a rigid fixed spheroid. Panels (A) and (B) correspond to a prolate (with an aspect ratio $a/b = 2$) and an oblate spheroid ($a/b = 1/2$), respectively. The spheroid is insonified by plane waves travelling along the $z$-direction. The solid lines are computed using the boundary matching method presented here, while the circles correspond to the numerical calculations obtained from the $T$-matrix approach, given previously in Fig. 4 (for a prolate spheroid) and Fig. 9 (for an oblate spheroid) of [25], respectively. For panel (B), $r_0$ in (13) equals $b$ so as to correlate the plot with Fig. 9 of [25]. As noticed from the curves, complete agreement is found between the results of the two formalisms.

and $P_n^1(\cos\theta) = -dP_n(\cos\theta)/d\theta$, are the associated Legendre functions of the first order. Note that the functions $\Gamma_n(\theta)$ and $\Lambda_n(\theta)$ depend only on the variable angle $\theta$ (for a fixed frequency or wave number $k$). To be able to solve the system of linear equations (6), the angular dependency must be eliminated. The procedure requires expanding the boundary condition equation (4) in PWSEs with separable variables, and matching each partial wave $n$. Accordingly, (6) is equated to a simplified Laplace series as,

$$\sum_{n=0}^{\infty} i^n (2n+1) P_n(\cos\beta) [\Gamma_n(\theta) + s_n \Lambda_n(\theta)]$$
$$= \frac{1}{2} \sum_{n=0}^{\infty} (2n+1) [\Upsilon_n + s_n \Omega_n] P_n(\cos\theta) \quad (8)$$
$$= 0,$$

where the term $(2n+1)/2$ is introduced for convenience, and $\Upsilon_n$ and $\Omega_n$ are the coefficients [independent of the angle $\theta$] to be determined after applying to (8) the orthogonality condition of the Legendre functions [45],

$$\frac{(2n+1)}{2} \int_0^\pi P_n(\cos\theta) P_{n'}(\cos\theta) \sin\theta \, d\theta = \delta_{n,n'}, \quad (9)$$

where $\delta_{n,n'}$ is the Kronecker delta function.

Equating the left- and right-hand sides in (8) for each partial wave, a new system of linear equations is obtained, which allows appropriate determination of the scattering coefficients $s_n$ for the spheroid. It is now rewritten as,

$$\sum_{n'=0}^{\infty} [\Upsilon_{n'} + s_n \Omega_{n'}] = 0, \quad (10)$$

where,

$$\begin{Bmatrix} \Upsilon_{n'} \\ \Omega_{n'} \end{Bmatrix} = \sum_{n=0}^{\infty} i^n (2n+1) P_n(\cos\beta)$$
$$\times \int_0^\pi \begin{Bmatrix} \Gamma_n(\theta) \\ \Lambda_n(\theta) \end{Bmatrix} P_{n'}(\cos\theta) \sin\theta \, d\theta. \quad (11)$$

Note that the integrals in (11) solely depend on the geometrical shape of the object, whereas the coefficients $\Upsilon_{n'}$ and $\Omega_{n'}$ are the result of the coupling between the coefficients characterizing the incident beam with the functions describing the geometric shape of the object.

The procedure for determining $s_n$ requires first the determination of (11) by numerical integration using (3). Once the scattering coefficients $s_n$ are obtained, they can be used to compute the scattered field via (2) on [i.e., $r = S(\theta)$] or near [$r \gtrsim S(\theta)$] the surface of the spheroid (corresponding to the near-field scattering), or far from its surface [$r \gg S(\theta)$], which corresponds to the far-field scattering.

It is common to evaluate the acoustic scattering in the far-field region. In that limit $(kr \to \infty)$, the steady-state (time-independent) scattered pressure field given in (2) can be expressed as,

$$P_s(r,\theta,\phi) \underset{kr\to\infty}{\approx} \frac{r_0 P_0}{2r} f_\infty(kr_0,\theta,\beta) e^{ikr}, \quad (12)$$

where $r_0 = max(a, b)$, and the spherical Hankel function of the first kind reduces to the following asymptotic approximation;

$$h_n^{(1)}(kr) \xrightarrow[kr\to\infty]{} \frac{1}{i^{(n+1)} kr} e^{ikr}.$$

The far-field form function is therefore expressed as,

$$f_\infty(kr_0,\theta,\beta) = \frac{2}{ikr_0} \sum_{n=0}^{\infty} (2n+1) s_n P_n(\cos\theta) P_n(\cos\beta).$$
$$\quad (13)$$

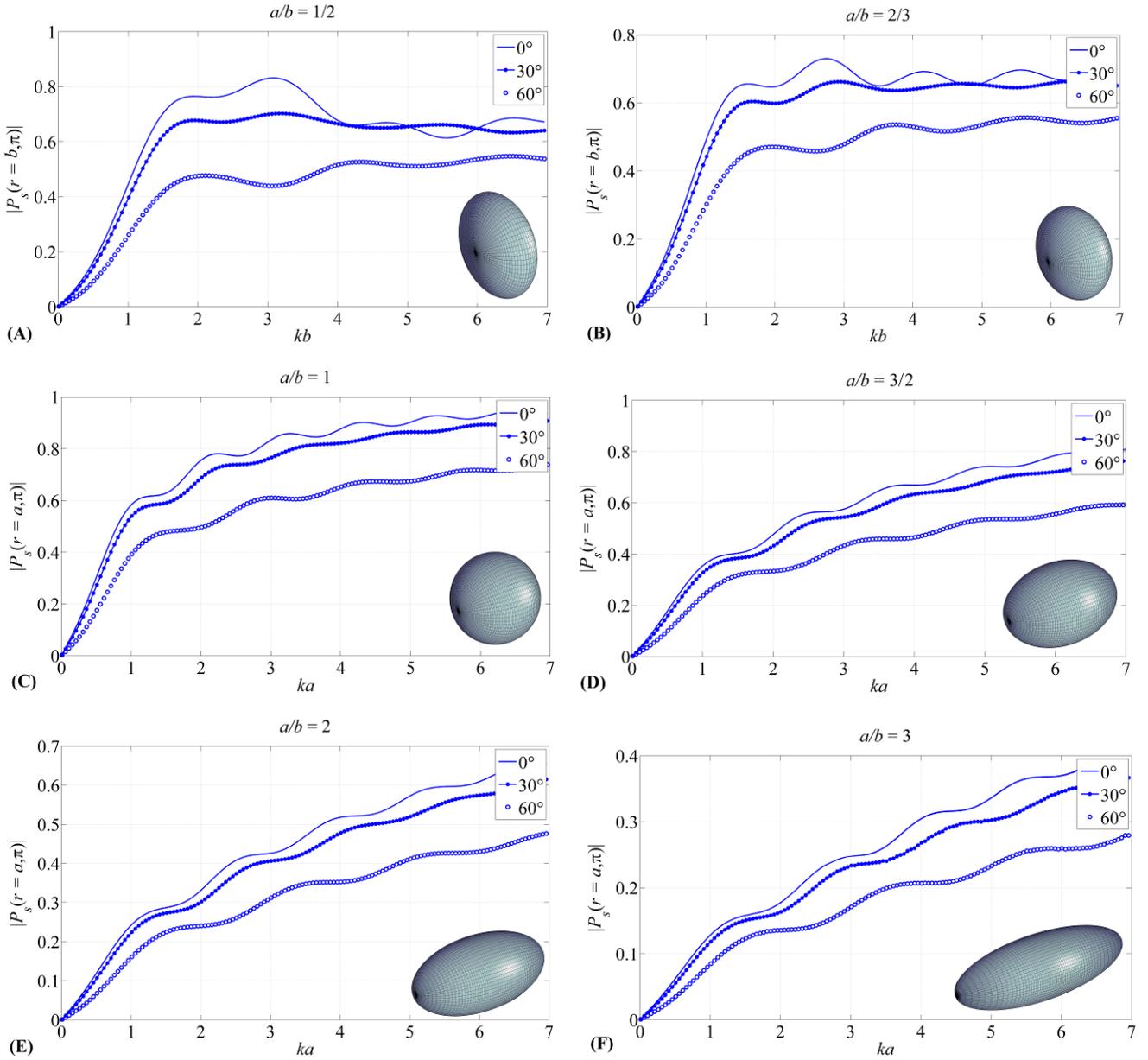

Fig. 3. The effect of varying the half-cone angle $\beta$ of the Bessel beam on the modulus plots of the end-on incidence backscattering ($\theta = \pi$) pressure in the near-field at $r=b$ for a rigid immovable (fixed) oblate spheroid [Panels (A), (B)], and at $r = a$ for a rigid fixed sphere [Panel (C)], and rigid fixed prolate spheroids [Panels (D)-(F)]. The ratio $a/b$ is given on the top of each figure. The case where $\beta = 0°$ corresponds to infinite plane waves.

## III. NUMERICAL RESULTS AND DISCUSSIONS

The coefficients $s_n$ are calculated by developing a MATLAB code to obtain the numerical solution of the system of linear equations (10). The integrals in (11) have been determined using a standard numerical integration procedure based on the fast trapezoidal method with a numerical sampling step of $\delta\theta = 10^{-5}$ that ensures convergence. Out of necessity, the series have to be truncated at some limit $N_{max}$, which corresponds to the maximum value of $n$ required for convergence. Ideally, the value of $N_{max}$ could be continually increased until a convergence criterion is satisfied leading to a negligible numerical error [46]. Practically, such convergence has been previously attained for quasi-spherical particles in electromagnetic scattering [47], nevertheless, the system of linear equations for highly elongated (typically a prolate surface with an aspect ratio larger than 3) or highly flat (disk-like oblate) geometries may become unsolvable because of an ill-conditioning during matrix inversion procedures. This instability arises from taking a large number of spherical partial-waves of order $n$ to fit a non-spherical object [48]. This also applies for the case where the frequency of the incident beam increases, as more terms in the series are needed to ensure convergence. With the increasing number of terms, the computational accuracy of the spherical Hankel functions begins to degrade rapidly, and creates the ill-conditioning [49]. This ill-conditioning should not be interpreted as a lack of rigor of the PWSE method [47], rather this technique may not be suitable from a computational standpoint for the analysis of highly elongated or extremely flat objects. For the purpose of the present study, the numerical analyses have been mainly concerned with a spheroid with a maximum aspect ratio of 3:1. In that limit, the proposed PWSE method provides adequate means to evaluate the scattering. The truncation order $N_{max}$ for the index $n$ in the series has been chosen such that $\left| s_{n+N_{max}} / s_0 \right| \sim 10^{-5}$, for $n = 1, 2,...$, which yielded a negligible truncation error.

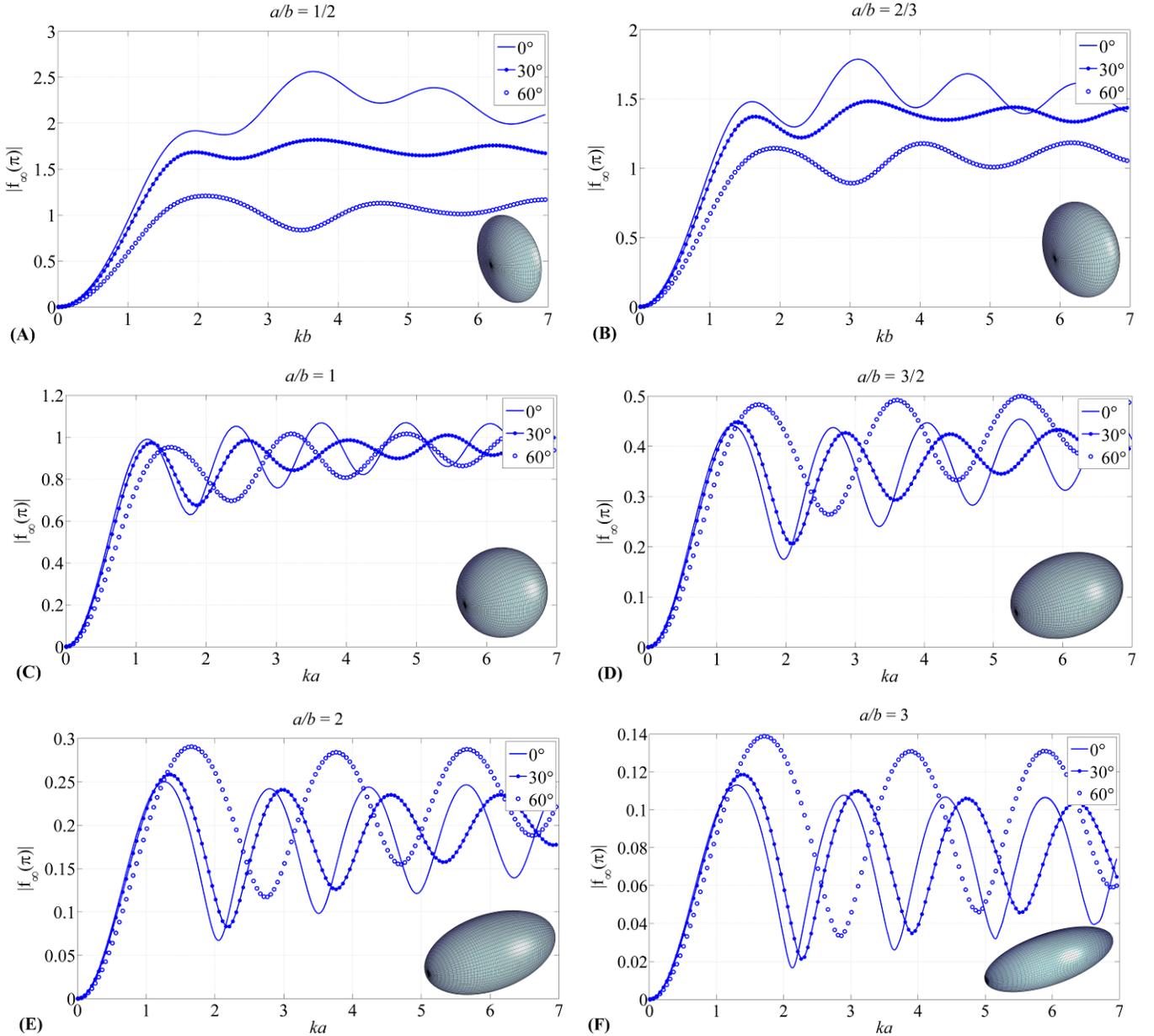

Fig. 4. The same as in Fig. 3, but the plots correspond to the end-on incidence backscattering form-function in the far-field ($kr \to \infty$).

*A. Numerical Validation*

The analysis is started by testing the validity of the formal solution based on the PWSE method with previous results (based on the *T*-matrix approach [25]) obtained for the far-field scattering of infinite plane waves propagating along the *z*-direction by a rigid spheroid. Panels (A) and (B) of Fig. 2 correspond to the magnitude of the far-field backscattering ($\theta = \pi$) form function given by (13) for a prolate (with an aspect ratio $a/b = 2$) and an oblate spheroid ($a/b = 1/2$), respectively, with a sampling step for the dimensionless frequency $\delta(kr_0) = 10^{-2}$. For the *oblate* spheroid case, the plots for the far-field backscattering form function are obtained for $r_0 = b$ in the denominator of (13), so as to correlate the results with those obtained previously [25]. The solid lines are computed using the formal solution of the PWSE method presented here, while the circles correspond to the numerical calculations obtained from the *T*-matrix approach, given in Fig. 4 (for the prolate spheroid case) and in Fig. 9 (for the oblate spheroid case) of [25], respectively. As noticed from the plots, complete agreement is found between the results of the two formalisms. Furthermore, additional tests and comparisons (not shown here for brevity) have been performed, and the results for the magnitude of the far-field backscattering form function using the formal solution of the PWSE method correlated exactly with those obtained using the shape perturbation method (introduced initially in electromagnetic scattering theory [50, 51]) presented in Fig. 9 of [36]. This effective mutual verification of the results with previous data obtained independently by different methods demonstrates the adequate validation of the formal solution using the PSWE method developed here.

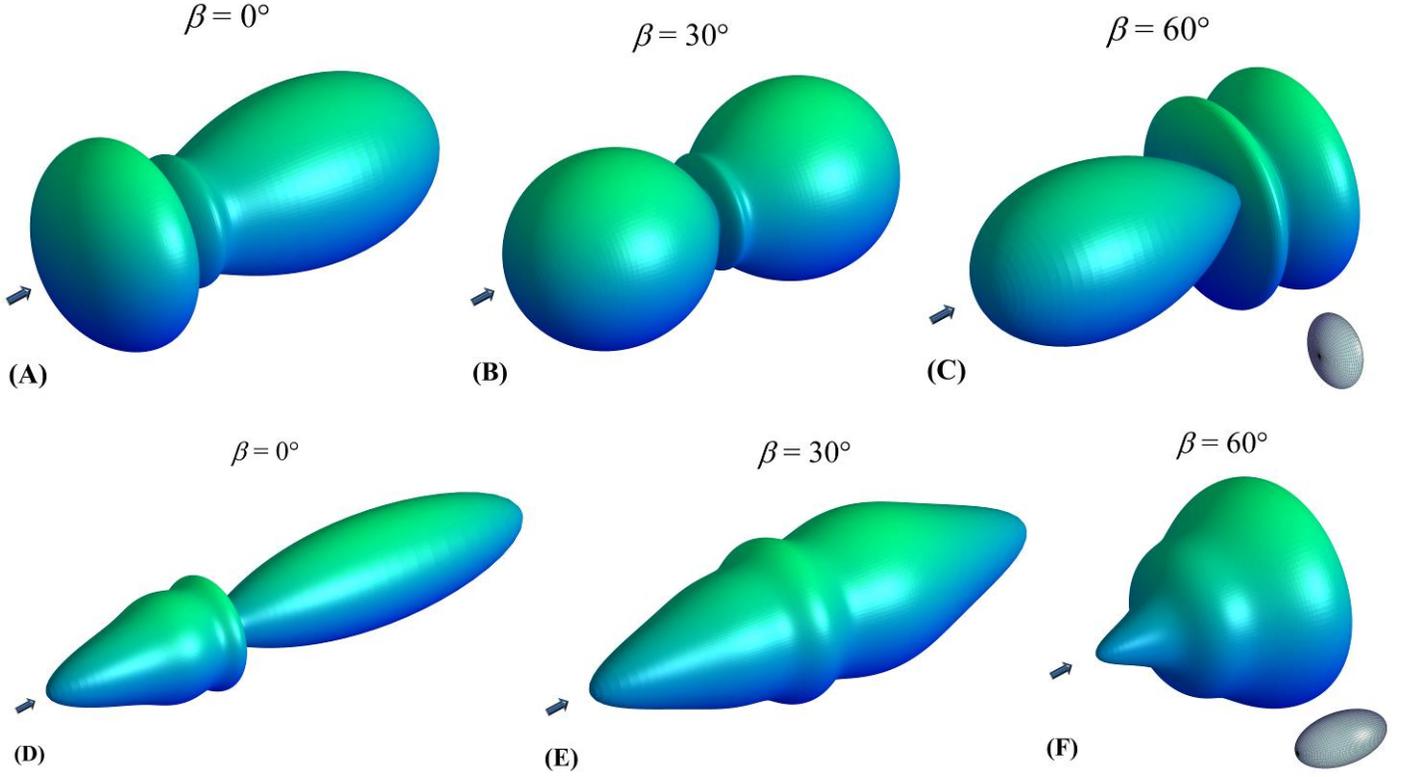

Fig. 5. Panels (A)-(C) display the 3D near-field scattering directivity patterns with end-on incidence at $r = b$ for a rigid oblate spheroid with $a/b = 1/2$, for $kb = 5$ and half-cone angles $\beta = 0°$, $30°$ and $60°$, respectively. Panels (D)-(F) correspond to the 3D near-field scattering directivity patterns at $r = a$ for a rigid prolate spheroid with $a/b = 2$, for $ka = 5$ and half-cone angles of the Bessel beam corresponding, respectively, to $\beta = 0°$, $30°$ and $60°$. The arrows on the left-hand side of each panel indicate the direction of the incident Bessel beam.

*B. Computations of the near- and far-field scattering from spheroids*

The analysis is illustrated by considering numerical computations for the backscattering pressure in the near-field and the backscattering form-function in the far-field, for spheroids ranging from oblate to prolate with end-on incidence.

Panels (A)-(F) in Fig. 3 display the transition from the oblate to prolate spheroid by changing the ratio $a/b$, for three values of the half-cone angle $\beta$ of the Bessel beam [$\beta = 0°$ (solid curves corresponding to plane waves), $\beta = 30°$ (solid-dotted curve), and $\beta = 60°$ (circles)]. The plots for the oblate spheroids ($a < b$) are computed versus $kb$ to facilitate the comparison with previous results [25]. As observed from these plots, it can be generally concluded that the backscattering pressure in the near-field is *reduced* as $\beta$ increases. Moreover, the transition from the sphere case [panel (C)] to the prolate cases [panels (D)-(F)] also reveals a reduction in the backscattering pressure magnitude.

In contrast with Fig. 3, panels (A)-(F) in Fig. 4 display the modulus of the backscattering form-function in the far-field versus $ka$ to also facilitate the comparison with previous results for plane waves [25], for the same half-cone angles used for Fig. 3. Unlike the near-field case, the far-field backscattering form-function magnitude increases as $\beta$ increases for the prolate spheroids [panels (D)-(F)]. This observation is in agreement with preliminary results obtained using the T-matrix formalism (See Fig. 2 in [29]). This may be explained as follows; in the near-field scattering from the spheroid, the contribution to the backscattering comes mainly from the individual partial-wave with an angle of incidence $\beta_n = 0°$ only (where the subscript $n$ denotes the $n^{th}$-partial wave) and backscattered in the direction $\theta = \pi$, whereas in the far-field, part of all the incident individual partial-waves with half-cone angles ranging between $0° \leq \beta_n \leq \beta$ backscatter in the direction $\theta = \pi$ and interfere constructively so as to produce the backscattering increase in the magnitude plots, the more prolate the spheroid becomes. Furthermore, the transition from the oblate case [panel (A)] to the prolate case [panel (F)] also reveals a reduction in the far-field backscattering form-function magnitude.

It is also important to note the periodic oscillations in the plots that are more pronounced in the far-field and as the dimensionless frequency $ka$ or $kb$ increases (Fig. 4). These oscillations are the result of the end-on incident (Bessel) waves that are specularly reflected from the edge of the spheroid, which interfere with slow circumferential waves (known as Franz' waves) propagating in the exterior fluid surrounding the spheroid [52]. As the spheroid's ratio $a/b$ increases, the frequency of these oscillations changes, resulting in a shift versus the dimensionless frequency. This shift is the result of a modification in the path length for the Franz waves circumnavigating the spheroid [29], and a change in their phase velocity.

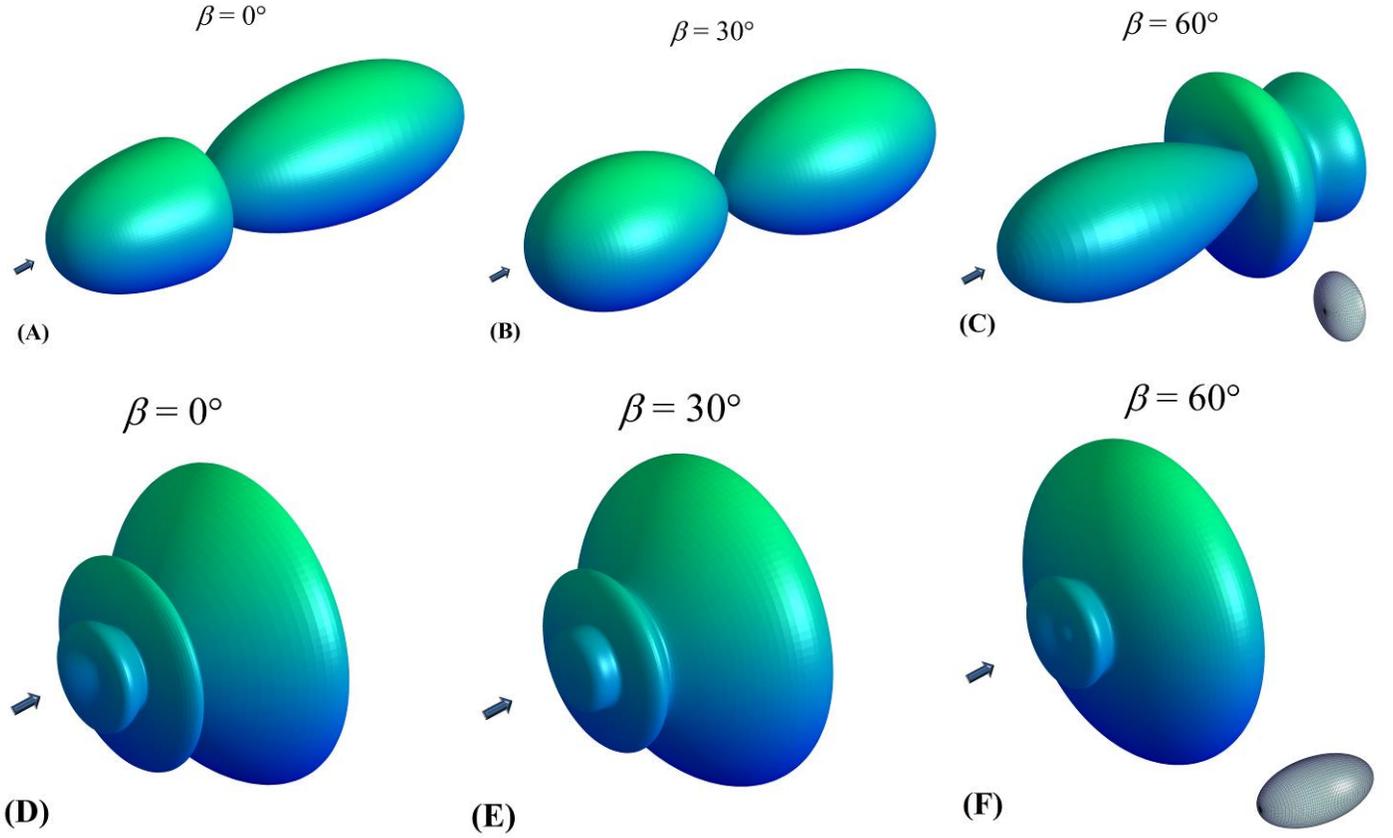

Fig. 6. Panels (A)-(C) display the 3D far-field scattering directivity patterns with end-on incidence [computed using (13)] for a rigid oblate spheroid with $a/b = 1/2$, for $kb = 5$ and half-cone angles $\beta = 0°$, $30°$ and $60°$, respectively. Panels (D)-(F) correspond to the 3D far-field scattering directivity patterns [computed using (13)] for a rigid prolate spheroid with $a/b = 2$, for $ka = 5$ and half-cone angles of the Bessel beam corresponding, respectively, to $\beta = 0°$, $30°$ and $60°$.

To further illustrate the difference between the scattering in the near- and far-field from the spheroids, the 3D directivity patterns are computed for the same half-cone angle values selected previously. Panels (A)-(C) in Fig. 5 display the 3D near-field scattering directivity patterns with end-on incidence at $r = b$ for a rigid oblate spheroid with $a/b = 1/2$, for $kb = 5$, and half-cone angles $\beta = 0°$, $30°$ and $60°$, respectively. Panels (D)-(F) in Fig. 5 correspond to the 3D near-field scattering directivity patterns at $r = a$ for a rigid prolate spheroid with $a/b = 2$, for $ka = 5$ and half-cone angles of the Bessel beam corresponding, respectively, to $\beta = 0°$, $30°$ and $60°$. Comparison of the 3D scattering directivity patterns for the oblate and prolate spheroids reveal significant differences, with particular emphasis on the choice of $\beta$. For example, in panel (C) of Fig. 5, an enhancement in the near-field pressure magnitude for the oblate spheroid case is manifested by a large lobe in the backscattering direction for $\beta = 60°$. This enhancement is less pronounced for the prolate spheroid case [panel (F)], which is associated with a reduction of the scattering in the forward ($\theta = 0°$) direction.

The far-field results are shown in Fig. 6. Panels (A)-(C) display the 3D far-field scattering directivity patterns with end-on incidence [computed using (13)] for a rigid oblate spheroid with $a/b = 1/2$, for $kb = 5$ and half-cone angles $\beta = 0°$, $30°$ and $60°$, respectively. Panels (D)-(F) correspond to the 3D far-field scattering directivity patterns [computed using (13)] for a rigid prolate spheroid with $a/b = 2$, for $ka = 5$ and half-cone angles of the Bessel beam corresponding, respectively, to $\beta = 0°$, $30°$ and $60°$. The backscattering enhancement observed initially in panel (C) of Fig. 5 for the near-field case, remains manifested in the far-field as shown in panel (C) of Fig. 6 for the oblate spheroid. Nevertheless, the difference in the 3D scattering directivity patterns between the near-field [panels (D)-(F) of Fig. 5] and the far-field results [panels (D)-(F) of Fig. 6] is significantly more distinct. Moreover, panels (A) and (B) in Fig. 6 for the oblate spheroid display a dipole-like radiating pattern, which may result from the direct contribution of the dipole ($n = 1$) partial-wave to the scattering.

Additional computations are performed for the far-field scattering form function [given by (13)] of rigid spheroids in the field of Bessel waves, with particular emphasis on their shapes (i.e. oblate vs. prolate by varying $a/b$), the scattering angle $\theta$ and the half-cone angle $\beta$ at a fixed non-dimensional frequency $ka$. The panels in Fig. 7 display the form function moduli plots in the bandwidths $0° \leq \theta \leq 180°$ and $0° \leq \beta < 90°$, where the rows (A)-(F) correspond to the transition from the oblate to the prolate spheroid geometry such that $a/b = 1/2$; 2/3; 1; 3/2; 2 and 3, respectively. The three columns correspond to computations with three different non-correspond to computations with three different non-dimensional frequencies, namely, $ka = 7$; 15 and 25, respectively. In all cases, the spheroids are centered on the axis of wave propagation of the Bessel waves. The most distinctive feature in the plots is the scattering enhancement starting at $\theta = \beta = 0°$, varying linearly across the plots (for the most panels), and ending at $\theta = \beta = 90°$. Meanwhile, the scattering enhancement narrows for increasing $\beta$; this result is expected since the beam's diameter at the first radial pressure node (i.e., central maximum) is proportional to $(1/\beta)$ [53]. For the oblate spheroid and sphere cases [i.e. panels (A)-(C)], the angles for the scattering enhancement always occur at $\theta \approx \beta$ for larger $ka$. Note that the maximum enhancement occurs at lower angle $\theta$ (or $\beta$) *values* that further decrease as $ka$ increases. For the plane wave case ($\beta = 0°$), it is known from the standard scattering theory for a sphere [52, 54] that the forward scattering magnitude (in the forward direction $\theta = 0°$) is the strongest in an angular scattering plot (See also Fig. 7-(C) for $\beta = 0°$). Since the incident Bessel waves constitute a superposition of plane waves propagating over a cone with a half-angle $\beta$, it is then anticipated that the scattering would be large at $\theta \approx \beta$, which corresponds to the forward scattering of the individual $n^{th}$ plane-wave component with half-cone angle $\beta_n$. For instance, recent investigations in optical scattering of Bessel waves from a *sphere* revealed a similar behavior [55-58].

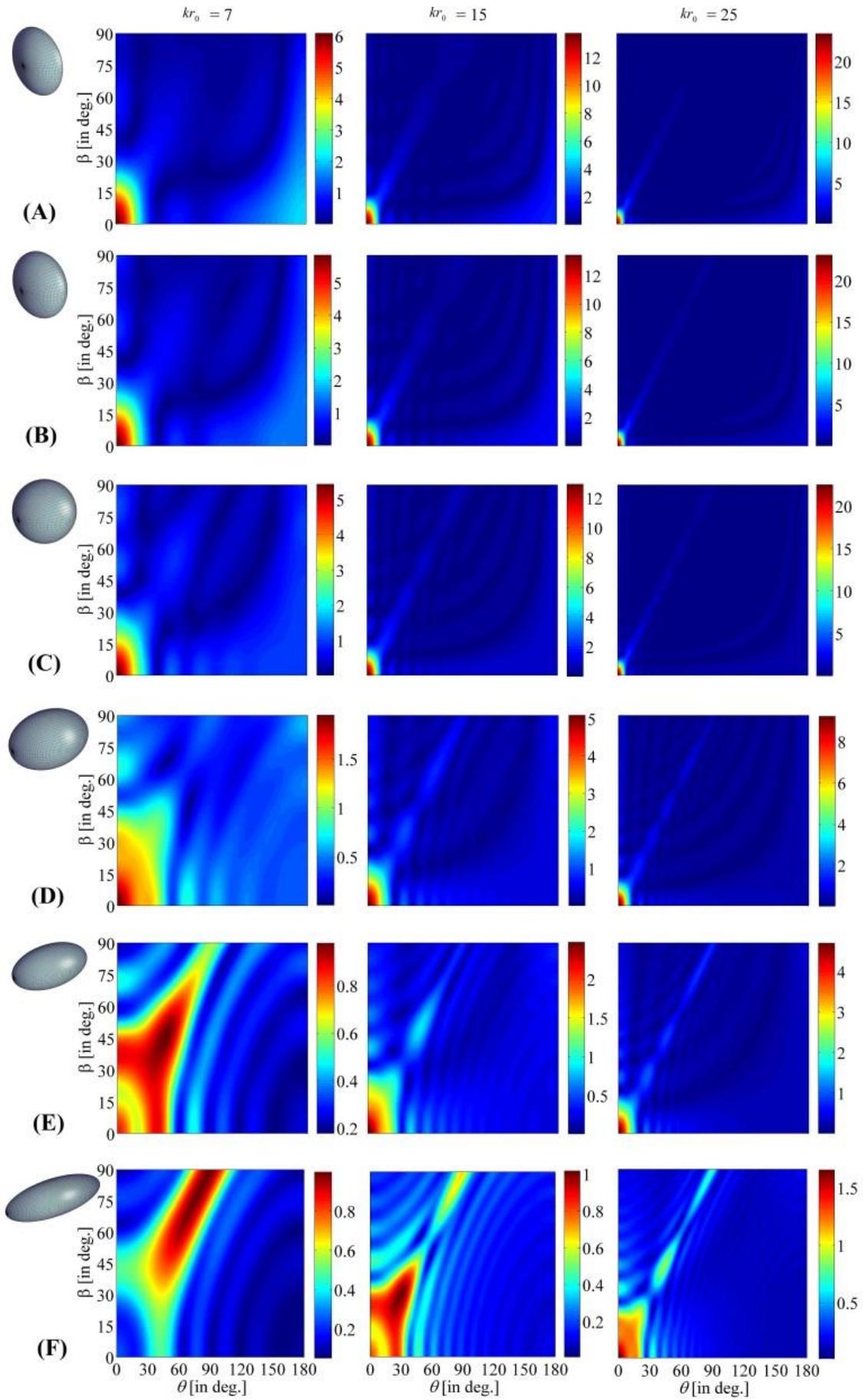

Fig. 7. The moduli plots for the far-field scattering form function [given by (13)] of rigid spheroids in the field of Bessel waves, with particular emphasis on their shapes (i.e. oblate vs. prolate by varying $a/b$), the scattering angle $\theta$ and the half-cone angle $\beta$ at three fixed non-dimensional frequencies $kr_0$ [where $r_0 = max(a,b)$] = 7; 15 and 25, respectively. Rows (A)-(F) show the transition from the oblate to the prolate spheroid geometry such that $a/b$ = 1/2; 2/3; 1; 3/2; 2 and 3, respectively.

Significant changes appear in the plots the more prolate the spheroid becomes, as shown in panels (D)-(E); at $ka = 7$, the scattering enhancement is extended over a large region in ($\theta$, $\beta$), for which its values decrease as $ka$ increases. Furthermore, one notices the presence of several arc-shaped undulations/ripples (i.e., scattering side-lobes) with variable periodicities with maxima and minima in ($\theta$, $\beta$), occurring on either side of the maximum scattering enhancement. More scattering side-lobes tend to appear as $ka$ increases, and a curving in the plots is noticed as $\beta$ becomes larger. The scattering side-lobes are caused by acoustic wave diffraction and the interference of surface waves matching their phases after repeated circumnavigations around the rigid (sound impenetrable) scatterer so as to create standing waves in the surrounding fluid.

## IV. CONCLUSION

This paper presents a PWSE method for predicting numerically the acoustical scattering from rigid (sound impenetrable) immovable oblate and prolate spheroids using a formal solution based on the separation of variables in spherical coordinates, assuming end-on (axial) incidence of a zeroth-order Bessel beam. The main advantage of the proposed method is the analogy with the partial-wave representation of the scattering by a spherical particle, which provides an important analytical tool to characterize the scattering by spheroids in spherical coordinates. Appropriate PWSEs for the incident and scattered pressure fields are derived, and numerical computations illustrate the analysis with particular emphasis on the half-cone angle $\beta$ of the Bessel beam, the dimensionless frequency, as well as the ratio $a/b$ (= major axis/minor axis) of the spheroid. The benchmark analytical solution could be used to validate results obtained by strictly numerical methods using the FEM, the FDTDM or other numerical tools. The method for predicting the acoustic scattering from elongated or flat objects finds important applications in various fields in underwater acoustics, particle levitation and radiation force [19], sizing and target identification, and especially when the physical phenomena under investigation involve the interaction of a rigid spheroid with acoustical Bessel waves.